\documentclass[useAMS, usenatbib]{mn2e}
\usepackage{graphicx}
\usepackage{txfonts, ulem}

 \title[Quark nova imprint in SN 2006gy]{Quark nova imprint in the extreme supernova explosion SN 2006gy}
   \author[R. Ouyed et al.]{R. Ouyed,$^1$\thanks{email:rouyed@ucalgary.ca} M. Kostka,$^1$ N. Koning,$^1$ D.A. Leahy,$^1$ and W. Steffen$^2$\\   
   $^1$ Department of Physics and Astronomy, University of Calgary,2500 University Drive NW, Calgary, Alberta, T2N 1N4 Canada \\
   $^2$ Instituto de Astronom\'{i}a Universidad Nacional Autonoma de M\'{e}xico, Ensenada, B.C., Mexico
          }
\begin{document}
  
 \maketitle

  \begin{abstract}
    The extremely luminous supernova 2006gy (SN 2006gy) is among the most energetic ever observed. The peak brightness was 100 times that of a typical supernova and it spent an unheard of 250 days at magnitude -19 or brighter. Efforts to describe SN 2006gy have pushed the boundaries of current supernova theory.
   In this work we aspire to simultaneously reproduce the photometric and spectroscopic observations of SN 2006gy using a quark nova model.
   This analysis considers the supernova explosion of a massive star followed days later by the quark nova detonation of a neutron star.  We lay out a detailed model of the interaction between the supernova envelope and the quark nova ejecta paying special attention to a mixing region which forms at the inner edge of the supernova envelope. This model is then fit to photometric and spectroscopic observations of SN 2006gy.
   This QN model naturally describes several features of SN 2006gy including the late stage light curve plateau, the broad H$\alpha$ line and the peculiar blue H$\alpha$ absorption.  We find that a progenitor mass between $20 M_{\odot}$ and $40 M_{\odot}$ provides ample energy to power SN 2006gy in the context of a QN.   	
  
  
\end{abstract}


%
\section{Introduction}
\label{intro}  

The supernova (SN) 2006gy discovered by Robert Quimby in August 2006 has challenged our understanding of stellar evolution \citep{quimby07}.  SN 2006gy was 100 times more luminous than a typical SN and at the time the most energetic ever recorded.  For almost a year it continued to radiate at a pace in which an ordinary SN could only sustain for at most a few days. Explaining this tremendous energy budget pushes the limits of current SN theory.

SN 2006gy reveals many singular features in both its light curve (left panel of Fig. \ref{obs}) and Hydrogen spectrum (right panel of Fig. \ref{obs}).  The light curve exhibits a luminous peak, broad shape and energetic plateau while the Hydrogen spectrum displays a curious evolution and a peculiar structure. Models have been proposed to describe individual characteristics of this event but all have been left wanting.

 \subsection{Supernova Explosion Models}  
 \label{SNmodels}
 The significance of SN 2006gy was first discussed independently by \cite{smith07} and \cite{ofek07}.  The model proposed by \cite{ofek07} involved a type Ia SN exploding during the common envelope phase of a binary system.  The collision between the SN ejecta (SNE) and dense circum-stellar material (CSM) releases the amount of energy required to explain the observations of SN 2006gy.  However this model demands the CSM to be extraordinarily massive \citep{ofek07} suggesting a mass loss rate several orders of magnitude greater than expected \citep{yungelson08}.  In addition the spectrum of SN 2006gy indicates the presence of elements not seen in a type Ia SN \citep{smith10}.

\cite{smith10} championed a CSM model which considered a wind from an exotic luminous blue variable (LBV) to account for the massive CSM.  Although this model can account for the first five months of the observed light curve, the diffusion process requires a rapid drop off in luminosity \citep{agnoletto09} rather than the observed plateau.  Attempts to reconcile the CSM model with the late stage plateau invoke the decay of radioactive $^{56}$Ni  and $^{56}$Co to generate the needed luminosity.  This necessitates 10 to 27 times the maximum amount of $^{56}$Ni that can be created by a SN \citep{umeda08}.    This description for the late stage light curve of SN 2006gy has been rebuked by recent near-infrared observations which show that the decline in luminosity is inconsistent with $^{56}$Co decay \citep{miller10}.  A further challenge to the CSM model lies in the fact that a high mass-loss rate is deduced using an equation that requires the stellar wind to be constant in time; however the conclusion made by \cite{smith10} is that the stellar wind must vary with time, contradicting their initial assumptions \citep{dwarkadas11}.  For this model to achieve the amount of radiated energy observed in SN 2006gy, the initial kinetic energy of the SNE must be at least $5\times10^{51}$ erg.  This implies that SN 2006gy was one of the most energetic SNe and requires a massive ($M_{\rm CSM} > 20 M_{\odot}$) CSM cloud \citep{smith10}. As noted by \cite{smith10}, the transformation of kinetic energy into radiation should translate into substantial narrowing of the H$\alpha$ line as the fast moving ejecta slows, which is contrary to that observed.  Finally any model that involves the collision of SNE with a dense CSM should be a strong emitter of X-ray radiation since the shock temperature would be very high \citep{blinnikov08}.  Observations from the \textit{Chandra X-ray observatory} of SN 2006gy have seen very limited X-ray emission \citep{smith10}.   The CSM model explains the lack of X-rays through self-absorption by the cold, outer layers of the CSM \citep{blinnikov08}.

A pulsational pair-instability (pPISN) model for SN 2006gy was considered by \cite{woosley07}.  In this model an unusually massive star ( $> 100$ M$_{\odot}$) becomes prone to the $\gamma = 4/3$ instability, triggering a SN  explosion.  The pair-instability process must occur twice, leading to a collision between ejected shells which releases the energy needed for SN 2006gy.  In order to reach the peak luminosity, it was necessary for \cite{woosley07} to artificially increase the kinetic energy of the second ejection.  Like the CSM model, the pPISN model fails to achieve on several levels.  First it is unable to properly account for the light curve plateau, as it falls off too rapidly \citep{woosley07}.  \cite{woosley07} explain that the extra energy required to fit the plateau must by generated by radioactive decay of $^{56}$Co, a conclusion contradicted by late stage near-infrared observations \citep{miller10}. Secondly, the multi-component structure of the H$\alpha$ line is difficult to reconcile with the pPISN model which demands all the Hydrogen to be contained in an outer shell \citep{woosley07, blinnikov08}.
Thirdly, a progenitor star of 110 $M_{\odot}$ is required for the pPISN model to achieve the power output of SN 2006gy.  A star this massive is expected to lose its Hydrogen long before it goes SN \citep{yungelson08}.  In order for numerical simulations to create a massive star with sufficient Hydrogen the merger of several young high mass stars was required \citep{yungelson08}.  This type of run-away stellar merger would be common in the early Universe but atypical 240 million years ago \citep{yungelson08}, at the era of the SN 2006gy explosion.    

  In this paper we present the quark nova as an alternative explanation for SN 2006gy.  Section 2 describes the physics of the interaction between the supernova envelope and quark nova ejecta.  In section 3, the photometric and spectroscopic fits of the quark nova model to observations of SN 2006gy are presented.  Section 4 provides a discussion of other observations of SN 2006gy which are consistent with the quark nova model.  Finally a conclusion is presented section 5 and future work discussed in section 6.    

 \section{Quark Nova Model}
 \label{QNmodel}
The quark nova (QN) was proposed as an  alternative explanation for SN 2006gy \citep{leahy08, ouyed09}.  A QN is expected to occur when the core density of a neutron star reaches the quark de-confinement density and triggers a violent \citep{ouyed02} conversion to the more stable strange quark matter \citep{itoh70,bodmer71,witten84}.  The novel proposition was made that during the spin-down evolution of the neutron star a detonative \citep{nieber10,ouyed10} phase transition to up-down-strange triplets would eject the outer layers of the neutron star at ultra-relativistic velocities \citep{keranen05,ouyed09a} (see the first panel of Fig. \ref{schematic}).  Follow-up studies of neutrino and photon emission processes during the QN \citep{vogt04,ouyed09b} have shown that these outermost layers ($10^{-4}-10^{-3} M_{\odot}$) can be ejected with up to $10^{53}$ erg in kinetic energy.  Nucleosynthesis simulations of the evolution of the neutron-rich QN ejecta (QNE) was found to produce primarily heavy elements with mass number, $ A> 130$ \citep{jaikumar07}.  
   
If the time ($t_{\mathrm{QN}}$) between SN and QN explosions is lengthy the SNE will have dissipated such that the QN essentially erupts in isolation.  However, when $t_{\mathrm{QN}}$ is on the order of days a violent collision occurs reheating the extended SNE to a temperature of $\sim10^{9}$ K \citep{leahy08,ouyed09}.  The brilliant radiance of the re-shocked SNE (RSNE) fades as the photosphere recedes, eventually revealing a mixture of the inner RSNE and the deposited QNE material.  In this work we revisit and extend that of \cite{leahy08} by introducing effects from the QNE to both the light curve and spectra of SN 2006gy.
    
\subsection{The Re-shocked Supernova Envelope}
\label{RSNEsect}
The collision between the QNE and SNE creates a shock that propagates at a speed $v_{\mathrm{shock}}$ through the SNE (see the middle panel of Fig. \ref{schematic}). The shock breaks out of the SNE at a distance $R_{\mathrm{RSNE,0}}$ at time 
\begin{center}
\begin{equation}
 t_{\mathrm{SBO}} =~t_{\mathrm{QN}}~+~ \frac{R_{\mathrm{RSNE,0}}}{v_{\mathrm{shock}}}. 
\end{equation}
\end{center}
  The RSNE which has a mass $M_{\mathrm{RSNE}}$ is therefore fully re-shocked at a radius 
\begin{center}
\begin{equation}
  R_{\mathrm{RSNE,0}}\,= \frac{ R_{*}~+  ~\,v_{\mathrm{SN,max}}\,t_{\rm QN}}{1-\frac{v_{\mathrm{SN,max}}}{v_{\mathrm{shock}}}},
\end{equation}
\end{center}  
   where $R_{*}$ is the radius of the progenitor star and $v_{\mathrm{SN, max}}$ is the maximum velocity of the homologously ($v\,\propto\,r$) expanding SNE.  The outer edge of the RSNE grows as 

\begin{center}
\begin{equation}
 R_{\mathrm{RSNE}}(t)=~R_{\mathrm{RSNE,0}}~+~v_{\mathrm{SN,max}}\,\times\,\left(t- t_{\mathrm{SBO}}\right).
\end{equation}
\end{center}
 The core of the spherical RSNE will cool adiabatically from its initial shock temperature ($T_0$) as,
\begin{center}
\begin{equation}
T_{\mathrm{RSNE}}^{\rm core}(t)=T_0 ~ \left( \frac{R_{\mathrm{RSNE,0}} } {R_{\mathrm{RSNE}}(t) }\right)^2. 
\end{equation}
\end{center}  
 For this proof-of-principle analysis we consider the temperature to follow a power law profile as: $T_{\rm RSNE}(r,t) \propto T_{\mathrm{RSNE}}^{\rm core}(t) \times r^{-\beta}$, where $\beta$ is a constant.     
 
To ensure that energy is conserved as continuum radiation is emitted by the hot RSNE (hRSNE) an equivalent amount of thermal energy is removed.  Since the hRSNE is optically thick the radiative cooling process starts from the outer edge and a photosphere propagates inward leaving behind a cold outer layer (cRSNE), as shown in the right panel of Fig. \ref{schematic}.  All species within the cRSNE are neutral and thus efficient absorbers of high energy radiation.  While we leave a rigorous treatment of X-ray absorption for future work, modelling column density of the cRSNE (see Fig. \ref{lcfit}) yields an explanation for why \textit{CHANDRA} observations show suppressed X-ray production by SN 2006gy \citep{smith07}.   

\subsection{The Hot Plate}
\label{HPsect}

\subsubsection{Formation}
The QN explosion sends ejecta propelling through space at speeds approaching c.  The ejecta will eventually catch up to and collide with the SNE material and mix at the inner edge of the RSNE.  The resulting composite shell will be referred to as the hot plate (HP) which has a constant thickness $\Delta R_{\mathrm{HP}}$.   Conservation of momentum tells us the HP will coast once it has captured all the SNE material it is capable of sweeping up.  The final momentum of the HP will be equal to the momentum of the QNE plus that of all the SNE material (represented by a velocity space integral) swept into the HP:

\begin{center}
\begin{equation} \label{eq:mhpvhp}
M_{\rm HP}  v_{\rm HP} =  \frac{E_{\rm QNE}}{c} + \int_{0}^{v_{\rm HP}} v\frac{dM_{\rm SNE}}{dv} dv,
\end{equation}
\end{center}  

where $E_{\rm QN}$ is the energy of the QNE, $M_{\rm HP}$ is the mass of the HP and $v_{\rm HP}$ is the final coasting speed of the HP.  Assuming homologous expansion ($r\propto v$) and a constant density ($\rho = M_{\rm t}/(4/3 \pi R_{\rm max}^3)$) the mass interior to velocity v is:

\begin{center}
\begin{equation}\label{eq:msne}
 M_{\rm SNE}(v) = M_{\rm t}\frac{v^3}{v_{\rm SN, max}^3}
\end{equation}
\end{center} 

where $M_{\rm t}$ is the total mass of the progenitor and $v_{\rm SN,max}$ is the velocity at the outer edge of the SNE.  Substituting equation \ref{eq:msne} into \ref{eq:mhpvhp} and solving for the velocity of the HP gives:

\begin{center}
\begin{equation}\label{eq:vhp}
 v_{\rm HP} = \left(\frac{4v_{\rm SN,max}^3E_{\rm QN}}{cM_{\rm t}}\right)^{\frac{1}{4}}
\end{equation}
\end{center} 

and using equation \ref{eq:vhp} in \ref{eq:msne}, assuming $M_{\rm SNE}(v_{\rm HP})\gg M_{\rm QNE}$, we have:

\begin{center}
\begin{equation}\label{eq:mhp}
 M_{\rm HP} = \left(\frac{4E_{\rm QN}}{cv_{\rm SN, max}}\right)^{\frac{3}{4}}M_{\rm t}^{\frac{1}{4}}
\end{equation}
\end{center} 

Several factors such as gravitational braking, reverse shocks and mixing will lead to lower $v_{\rm HP}$ values than that given by equation \ref{eq:vhp}.  A lower limit for $v_{\rm HP}$ is given by the escape speed $v_{\rm HP,esc}$ at the initial radius of the HP, $R_{\rm HP,0}$.  The time, $t_{\rm coast}$, for the HP to form and start coasting can be estimated as follows:  

The time taken for the QNE to catch up to and sweep up the HP material is $(t_{\rm coast}-t_{\rm QN})$, so the radius of the HP at time $t_{\rm coast}$ can be expressed as:

\begin{center}
\begin{equation}\label{eq:rtcoast2}
R_{\rm HP}(t_{\rm coast}) = \alpha c (t_{\rm coast} - t_{\rm QN})
\end{equation}
\end{center} 

where $\alpha c$ represents an average speed of the QN ejecta during the catch-up and sweeping phases.  We explored a large range physical parameters of the QNE and found that typically the HP enters the coasting phase between $1.5\,t_{\rm QN}$ and $3\,t_{\rm QN}$, thus we chose as a fiducial value $t_{\rm coast} \sim 2\, t_{\rm QN}$ after which the radius of the HP is described by: 

\begin{center}
\begin{equation}
R_{\rm HP}(t) = R_{\rm HP,0} + v_{\rm HP}\times (t_{\rm QN}-t_{\rm coast}), 
\end{equation}
\end{center}  
  
\subsubsection{Cooling}

For an adiabatic gas we have:

\begin{center}
\begin{equation}\label{eq:adiabatic}
T_{0}V_{0}^{\gamma -1} = TV^{\gamma -1}
\end{equation}
\end{center}  

Since the HP is a coasting shell of constant thickness its volume is given by $4\pi r^2 \Delta R_{\mathrm{HP}}$.  Using $\gamma = \frac{5}{3}$ with equation \ref{eq:adiabatic} thus gives the temperature as a function of time:

\begin{center}
\begin{equation}
T(t) = T_{0}\left(\frac{R_{\rm HP,0}}{R_{\rm HP}(t)}\right)^{\frac{4}{3}}
\end{equation}
\end{center}  

Photon diffusion will become important in the cooling of the HP when the HP and the overlying envelope become sufficiently transparent.  We can approximate when this will happen ($t_{\rm crit}$) by setting the mean free path to a fraction of the envelope size:

\begin{center}
\begin{equation}\label{eq:mfp}
\lambda \approx \frac{1}{n_{e} \sigma_{TH}} \approx \alpha R_{\rm SNE}(t_{\rm crit})
\end{equation}
\end{center}  

Up to a radius of $\alpha R_{\rm SNE}$ we assume heating by the HP completely ionizes the hydrogen in the envelope at all times $t < t_{\rm crit}$.  Beyond this, the envelope may or may not be ionized depending on its temperature.  The radius of the envelope at time $t_{\rm crit}$ is given by $R_{\rm SNE}(t_{\rm crit})=v_{\rm max}t_{\rm crit}$ which leads to (using equation \ref{eq:mfp}):

\begin{center}
\begin{equation}\label{eq:tcrit}
t_{\rm crit}=\left(\frac{3M_{\rm env}\sigma_{TH}\alpha}{4\pi v_{\rm SN, max}^2\mu m_{H}}\right)^\frac{1}{2}
\end{equation}
\end{center}  

Using values of $v_{\rm SN, max}=4000 \rm {kms}^{-1}$, $\alpha=0.01$, $M_{\rm env} = 25M_{\odot}$ and $\mu=1.2$ we get $t_{\rm crit}\approx 200$ days.  Once $t_{\rm crit}$ is reached, diffusion will quickly cool the HP within $\sim 1$ day.  This calculation demonstrates that the HP is able to retain its heat for hundreds of days.  Recognition of the sustained high temperature of the HP is essential to deciphering the enigma of SN 2006gy as this naturally describes both the late stage plateau in the light curve and the unyielding broad component of the H$\alpha$ line.  

\subsubsection{Emission}
 
Radiation from the HP is treated as diffusion luminosity, given by: 
\begin{center}
\begin{equation}
L_{\rm HP}= 4\pi \,R_{\rm HP,p}^2 n_{\rm HP}\, c_{\rm v} \, \Delta T \frac{dD}{dt},
\end{equation}
\end{center}   
where $c_{\rm v}$ is the specific heat, $n_{\rm HP}$ is the number density of the HP,  $\Delta T$ represents the temperature difference at the interface between the HP and RSNE.  The HP photosphere ($R_{\rm HP,p}$) defines the emitting surface, which moves at the diffusion rate, $\frac{dD}{dt}$ \citep{leahy08}.  
 
\subsection{Radiative Transfer}
\label{rtransfer}
 
The geometry of our model is quite simple as it contains only three components; a thin inner shell (the HP) surrounded by very thick outer shells (the hRSNE and cRSNE).  For our analysis we use the astrophysical modelling software  SHAPE\footnote{www.astrosen.unam.mx/shape} \citep{steffen10} to build a geometric model of the RSNE and HP.  Alongside being able to quickly and accurately construct 3-D geometries, density and temperature profiles, SHAPE is able to perform fast radiative transfer calculations.  

Radiative transfer in SHAPE is accomplished through a ray-casting algorithm.  The aforementioned 3D model is subdivided into a $128^3$-cell grid, with each cell representing a volume (voxel) through which radiative transfer is performed.  Given a temperature and density distribution, SHAPE calculates emission and absorption coefficients for each voxel in the grid.  A ray is then cast from behind the grid to the observer, performing radiative transfer calculations along each voxel for each frequency band.  The emerging ray can then be used in the generation of the light curve and spectrum plots.  This allows us to simultaneously produce light curves and spectra which can then be compared to the observations of SN 2006gy.  For this analysis we use rays consisting of 150 bands ranging from $3\times10^{14}$Hz - $3\times10^{18}$Hz.  The plot of the light curve only uses R-band frequencies from $\sim3.76\times10^{14}$Hz - $5.76\times10^{14}$Hz, but the energy conservation requires the entire range.   For the spectra, we use 200 bands from $4.4118\times10^{14}$Hz - $4.6875\times10^{14}$Hz (spanning the frequency of H$\alpha$).

Unlike most radiative transfer codes, SHAPE does not calculate the temperature of each cell in the grid.  Instead, the temperature is an input in the form of an analytic distribution.  Although less accurate than a full radiative transfer treatment, the speed gain allows for an interactive parameter search as required by this work.  In summary, SHAPE answers the question ``Given a temperature and density, what is the intensity of the light emitted by an object?".

\subsection{Model Parameters}
\label{params}
The parameters used to construct the light curve and spectra of SN 2006gy can be divided into three categories: Constrained, Free Physical, and Free Radiative Transfer.

\subsubsection{Constrained Parameters}
 The following parameters that are used in our model are constrained by observations.

\begin{itemize}

\item[$\bullet$]  $M_{\rm t}$: the total mass ($M_{\rm RSNE} + M_{\rm HP}$) is constrained from above by $\sim40 M_{\odot}$ (anything heavier will likely form a black hole rather than a neutron star \citep{heger03}) and  $\sim20 M_{\odot}$ from below (anything less massive will not form a quark star \citep{staff06}).

\item[$\bullet$]  $v_{\mathrm{SN, max}}$: the outer edge velocity of the expanding RSNE is constrained to be 4000 km$\cdot$s$^{-1}$ by the blue-side absorption feature (as noted in Fig. \ref{d96}) observed in the H$\alpha$ spectrum \citep{smith10}.   

\item[$\bullet$]  $T_0$: the initial temperature of the RSNE and HP is constrained by fitting the broad component of the H$\alpha$ line for each spectrum assuming it is thermally broadened.  The extrapolated $T_0$ value is on the order of $\sim1\times$10$^9$K.

\item[$\bullet$]   $v_{\mathrm{shock}}$: the speed of the shock as it travels through the RSNE is $\sim$6000 km$\cdot$s$^{-1}$, this is found using shock physics as $k\,T_0 = \frac{3}{16}\mu\,m_H\,v_{\mathrm{shock}}^2$ where the initial temperature ($T_0$) is known.  

\end{itemize}

\subsubsection{Free Physical Parameters}

In our model the SN explosion of a massive star is followed a time $t_{\mathrm{QN}}$  later by a QN explosion.  The QNE/SNE collision sends a shock through the SNE creating two distinct layers: the HP and RSNE.  The HP has a mass $M_{\mathrm{HP}}$, coasting speed $v_{\mathrm{HP}}$, initial radius $R_{\mathrm{HP,0}}$ and constant thickness $\Delta R_{\mathrm{HP}}$.  The RSNE has a mass, $M_{\mathrm{RSNE}}$ and temperature profile defined by the power law index, $\beta$.  We assume constant density for all components.  These variables make up the free physical parameters in our model which we find by fitting our theoretical light curve to the observations of SN 2006gy.   

\subsubsection{Free Radiative Transfer Parameters}

The radiative transfer in SHAPE requires the specification of an absorption and emission coefficient for each frequency, $\nu$.  The emission coefficient used in this analysis is of the form:

\begin{center}
\begin{equation} \label{eq:j}
j_{\nu} \propto \frac{A n_e^2}{T^{3/2}} e^{h\nu/kT}
\end{equation}
\end{center}   

Where A is a multiplicative factor.  The corresponding absorption coefficient (assuming LTE) is:

\begin{center}
\begin{equation} \label{eq:kappa}
\kappa_{\nu} = \frac{j_{\nu}}{B_{\nu}}
\end{equation}
\end{center}   

Where $B_{\nu}$ is the Planck function.  The form of equation \ref{eq:j} is similar to that of bound-free continuum emission, perhaps hinting at an emission mechanism.  In addition to the absorption in equation \ref{eq:kappa}, we include a Thomson scattering term of the form:

\begin{center}
\begin{equation} \label{eq:kappa_th}
\kappa_{\nu, \rm TH} = B n_e \sigma_{\rm TH}
\end{equation}
\end{center}   

Where $\sigma_{\rm TH}$ is the Thomson scattering cross section and B is a multiplicative factor.  Due to the computational and time demands of treating multiple scattering, we simply assume that scattering is isotropic and everything removed from the beam will eventually be put back in.  Of course the scattering is not perfectly elastic and energy will be lost.  Therefore B represents the fraction of scattered light that will NOT be scattered back into the beam.

\section{Results}
\subsection{Photometry}
\label{photo}

Through the exploration of parameter space using SHAPE, we find that the light curve is well reproduced by a progenitor mass ranging from 20-40 $M_{\odot}$.  We have fit three models with progenitor masses of 20, 30 and 40 $M_{\odot}$ with the parameters of each fit listed in  table \ref{table:parameters}. Each uses an envelope temperature profile of $\beta=0.2$ with an initial temperature of $T_0=0.9\times 10^9 K$.  The initial energy of the QN ejecta is set to $\sim 1 \times 10^{52}$ ergs \citep{keranen05}  We find that the actual velocity of the HP differs from the theoretical (equation \ref{eq:vhp}) by a factor of $~0.1$.  As previously mentioned, factors such as gravitational braking, reverse shocks and mixing will lead to lower $v_{\rm HP}$ values.  

The emission is simulated using the coefficients of equations \ref{eq:j}, \ref{eq:kappa} and \ref{eq:kappa_th} with the A values listed in table \ref{table:parameters} and $B=5\times10^{-4}$. While the shock is still moving through the SNE ($t<t_{\rm SBO}$) detailed shock physics would be required to properly describe the corresponding light curve, thus we begin our photometric analysis at time $t=t_{\rm SBO} \simeq 30$ days. Our SN explosion date is 5, 10 and 20 days earlier than that inferred by \citet{smith10} for the 40, 30 and 20 $M_{\odot}$ fits respectively.

Figure \ref{figure:ms} shows the fits for the various progenitor masses, all of which match the data well.  Figure \ref{lcfit} shows how the different components contribute to the light curve.  The high initial temperature and radius of the RSNE provide ample luminosity (see the blue dash-dot line in Fig. \ref{lcfit}) to explain the broad peak of the observed light curve.  As the photosphere recedes, radiation from the underlying HP becomes more prominent and the drop in the overall light curve (red solid line in Fig. \ref{lcfit}) is mitigated by support from the HP (green dashed line in Fig. \ref{lcfit}).  The slowly coasting and thus slowly cooling HP provides a natural explanation for the plateau observed in the late stage of the SN 2006gy light curve.  

\subsection{Spectroscopy}
\label{spectro}

For our proof-of-principle spectroscopic analysis we aimed to determine the cause of the overall structure of the H$\alpha$ spectrum of SN 2006gy, as to date only a phenomenological analysis has been done (ie. fitting a Gaussian versus a Lorentzian profile \citep{smith10, chatzopoulos11}).  Displayed in Fig. \ref{HAfit} is our model H$\alpha$ line (red solid line) compared to spectral observations of SN 2006gy for days 36, 65, 71, 96, 125, 154, 179 and 209.  As shown in the right panel of Fig. \ref{schematic}, after shock break out our model contains three components; an outer layer of cold RSNE (cRSNE), a hot region of RSNE (hRSNE) and an innermost thin HP shell.  Each of these constituents contribute to the overall shape of the H$\alpha$ line.  For our modelling we ignore contributions from the two identified [N II] lines at 6548 and 6583 Angstroms.   

By viewing the width of the broad  H$\alpha$ component in the light of thermal rather than Doppler broadening, the reason for the curious late stage breadth becomes clear.  Considering the slowly coasting HP as the thermal source, the width of the broad component is then proportional to T$_{\rm HP}$.  Plotted with a green dash-dot line in Fig. \ref{d96} is the broad H$\alpha$ component for day 96 (width $\propto T_{\rm HP}=7.5\times 10^8$ K).  We determined the temperature evolution of the HP by fitting a Gaussian to the broad component of the H$\alpha$ line.  The decrease in temperature agrees remarkably well with the predicted HP temperature from our light curve (top-right panel of Fig. \ref{HPprop}).

The peculiar asymmetry of the H$\alpha$ line observed in SN 2006gy is inescapable in our scenario.  The wide blue-side absorption feature that does not dip below the continuum is due to RSNE material along our line of sight absorbing emission from the underlying HP.  The blue-side feature extends out 4000 km$\cdot$s$^{-1}$ as both the hRSNE and cRSNE contribute to the absorption.   The blue-side absorption is strongest at times prior to day 96 and then becomes weaker as the absorbing RSNE material becomes increasingly diffuse and increasingly cool.  The bottom right panel of Fig. \ref{HAfit} most clearly displays the evolution of the blue-side absorption in our model.  

At early times the narrow H$\alpha$ peak is broadened by thermal motion. At some time between day 71 and 96 the temperature of the hRSNE drops such that kinematic Doppler broadening takes over.  As the location of the H$\alpha$ emission region recedes into slower moving RSNE material, emission from the hRSNE originates from ever lower velocities.  This accounts for the discontinuity in the H$\alpha$ line at $- 1200$ km$\cdot$s$^{-1}$ on day 96 seen in Fig. \ref{d96}.  However the cold outer layer of the RSNE continues to absorb out to 4000 km$\cdot$s$^{-1}$ (see green dashed line in Fig. \ref{d96}).  

 The sharp P-Cygni profile located at the center of the H$\alpha$ line seen in observations on days 96, 125 and 154 is characteristic of an expanding (velocity $\approx 120$ km$\cdot$s$^{-1}$) shell of Hydrogen.  This velocity suggests the sharp P-Cygni feature is likely due to stellar wind from the progenitor star.  This feature does not seem to be typical for super-luminous SNe as it has not been observed in the super-luminous SNe: SN 2008am, \citep{chatzopoulos11}, SN 2008fz \citep{drake10}, SN 2003ma \citep{rest11}, SN 2008es \citep{miller09, gezari09} or SN 2007bi \citep{galyam10}. 
 
 For illustrative purposes we added an outermost shell of Hydrogen to our model in order to simulate a stellar wind ($\rho\, v \,r^2=$const.) from the progenitor.   Our low mass ($\approx 0.5$ M$_{\odot}$) stellar wind extends from the outer edge of the cRSNE to a radius of 3 $\times$ 10$^{14}$ m and has a temperature of 8000 K.  Fig. \ref{wind} displays a comparison of the observed evolution of the H$\alpha$ line of SN 2006gy with our (stellar wind included) QN model.  With the exception of the stellar wind, the same model parameters were used to create the fits seen in Figures \ref{HAfit} and \ref{wind}. As seen in Fig. \ref{wind} at early stages (days 36, 65 and 71) the stellar wind continuum is too low for absorption to occur and the stellar wind adds a narrow peak to the H$\alpha$ line. The continuum emission of the stellar wind then increases to a level that a strong P-Cygni profile becomes present by day 96.  The P-Cygni profile diminishes with time as the strength of the stellar wind continuum falls.  By day 209 the P-Cygni profile is gone leaving only a narrow emission peak as the stellar wind contribution to the H$\alpha$ line.  The bulk structure of the H$\alpha$ line can be described by emission from the HP and the RSNE (see fig. \ref{HAfit}), while the surrounding low mass stellar wind adds the fine details to the spectral line (see Fig. \ref{wind}).  A side by side comparison of our model H$\alpha$ line with the observed spectral line can be seen in Fig. \ref{Hoverlay}.

In this proof-of-principle analysis we found that the H$\alpha$ emission originated from a region interior to the photosphere (see Sect. 2.1 for definition of photosphere) at all times in the evolution of the spectrum.  This is to be expected as the photosphere defines the location of continuum emission while H$\alpha$ emission occurs at a chromospheric region which is dependent on the density and temperature of the plasma \citep{mihalas78}.  As future work, improvements must be made to our formalism to include such elements as asymmetries since higher order effects could have an impact on the spectroscopic behaviour of this system. A more detailed study of the structure of the RSNE and the role of the density and temperature profiles on the spectroscopy of the QN will as well be undertaken in future studies.  

\section{Discussion}
\label{discuss}

The HP temperature ($T_{\rm HP}$) provides us with an opportunity to check our model for consistency.  Comparing our model $T_{\rm HP}(t)= T_0 \left(\frac{R_{\rm HP,0}}{R_{\rm HP}(t)}\right)^{4/3}$ (red solid line in the top-right panel of Fig. \ref{HPprop}) with values found by fits to the H$\alpha$ line (black circles plotted in top-right panel of Fig. \ref{HPprop}) we can see that they are in good agreement. 

As noted by \cite{smith10} there exists a second peak of H$\alpha$ luminosity accompanied by peaks in Fe II and Ca II luminosity that evolve in concert (see the bottom panel of Fig. \ref{HPprop}). This is expected in our QN scenario, since during the creation of the HP, ploughing of the inner RSNE material will mix the elements leading to their simultaneous evolution.  As seen in the bottom panel of Fig. \ref{HPprop} the rise of the Fe II (red diamonds), Ca II (blue squares) and H$\alpha$ luminosity (black circles) correspond with the emergence of the HP (green dashed line).

During the formation of the HP, elements with $A>130$ (in the QNE, \cite{jaikumar07}) will mix with the lighter elements in the SNE.  We predict some $A>130$ elements should appear late in the evolution of the SN 2006gy spectrum evolving in concert with the Fe II and Ca II.  The intensity of the $A>130$ emission lines should be much lower than those from the RSNE as 
\begin{center}
\begin{equation}
I_{\rm A>130}\propto \frac{M_{\rm QNE}}{M_{\rm HP}} < 10^{-4}. 
\end{equation}
\end{center}   

Plotted as a solid black line in Fig. \ref{lcfit} is the column density of the cRSNE.  After shock breakout the column density of the neutral cRSNE rapidly rises to $10^{25}$ cm$^{-2}$, thus quickly becoming opaque to X-rays.  This naturally explains why \textit{CHANDRA} observations show minimal X-ray production from SN 2006gy.

We were able to reproduce the light curve of SN 2006gy using a range of progenitor masses.  The lower mass models required a higher emission coefficient ($A$ value) and longer time delays to recover the fit.  Because $T_0$ does not vary from one fit to another and the continuum levels remain the same (hence the same fit to the light curve), the spectra are unaffected by varying the above parameters. The range of possible $A$ coefficients highlights our ignorance of the emission process and more detailed radiative transfer calculations will be required to narrow down the progenitor mass and the time delay between SN and QN.  Regardless, our models do show that given a progenitor mass between $20M_{\odot}$ and $40M_{\odot}$, enough energy is available to explain the amplitude and breadth of the SN 2006gy light curve.

\section{Conclusion}
\label{concl}
 The QN model presented in this proof-of-principle analysis simultaneously provides a very good fit to both the light curve and spectra of SN 2006gy.  The broad, luminous peak of the light curve is described by radiation emitted from the hRSNE.  The late stage plateau in the light curve, which is thus far inexplicable in other models, is naturally explained by emission from the slowly coasting HP.  These distinctive features of the QN model also account for the peculiar structure and evolution of the H$\alpha$ line.  The persistent broad H$\alpha$ component, which is not understood in the context of any other model, is uniquely described by emission from the slowly cooling HP of the QN model.  The blue-side absorption feature observed in the H$\alpha$ line is caused by both the cRSNE and hRSNE which attenuate emission from the underlying HP.  As the hRSNE radiates energy it builds a cold outer layer (the cRSNE) which is opaque to high energy radiation.  This illustrates how an explosion as energetic as SN 2006gy can be surprisingly quiet in X-rays.  The QN is currently being used to model other super-luminous SNe.  The findings will be published in an upcoming paper.
 
\section{Future Work}
\label{future}
This work is intended to demonstrate that a QN explosion of a neutron star only days after the initial SN is able to account for the light curve and spectra of SN 2006gy.  However, the exact emission mechanism of the radiation is not well understood.  The emission coefficient that best reproduces the rise and fall of the light curve resembles bound-free continuum emission (boosted by a factor of A).  The boost may be due to the contribution of some other emission mechanism such as free-free.  Future work will concentrate on a more detailed treatment of radiative transfer than possible with the current computer code.

\section{Acknowledgements}
    RO, MK, NK and DAL are supported by the Natural Sciences and Engineering Research Council of Canada.  NK acknowledges support from Alberta Ingenuity and the Killam Trusts.  WS acknowledges support from grant UNAM PAPIIT IN100410.


\begin{onecolumn}

\begin{center}
\begin{table}
\caption{{\bf Model parameters for different progenitor masses ($M_{\rm t}$).}  Parameters listed with * are determined through equations and are not adjustable.}
	\begin{tabular}{|l|l|l|l|l|l|l|l|l}
	\hline
	$M_{\rm t}(M_{\odot})$ & $M_{\rm env}(M_{\odot})^{*}$ & $M_{\rm HP}(M_{\odot})^{*}$ & $t_{\rm QN}$ (days) & A & $R_{\rm HP, 0}$ (cm)$^{*}$ & $\Delta R_{\rm HP}$ (cm) & $v_{\rm HP}$ (kms$^{-1}$)$^{*}$  \\
	\hline
	40 & 36.3 & 3.7 & 10.5 & 1750 & $3.3\times 10^{14}$ & $2.5\times 10^{12}$ & 1809.7 \\
	30 & 26.6 & 3.4 & 11.5 & 4000 & $3.9\times 10^{14}$ & $5.0\times 10^{12}$ & 1944.7 \\
	20 & 16.9 & 3.1 & 12.5 & 12000 & $4.6\times 10^{14}$ & $10.0\times 10^{12}$ & 2152.1 \\
	\hline	
	\end{tabular}
	\label{table:parameters}
\end{table}
\end{center}

\begin{figure}
\resizebox{\hsize}{!}{\includegraphics{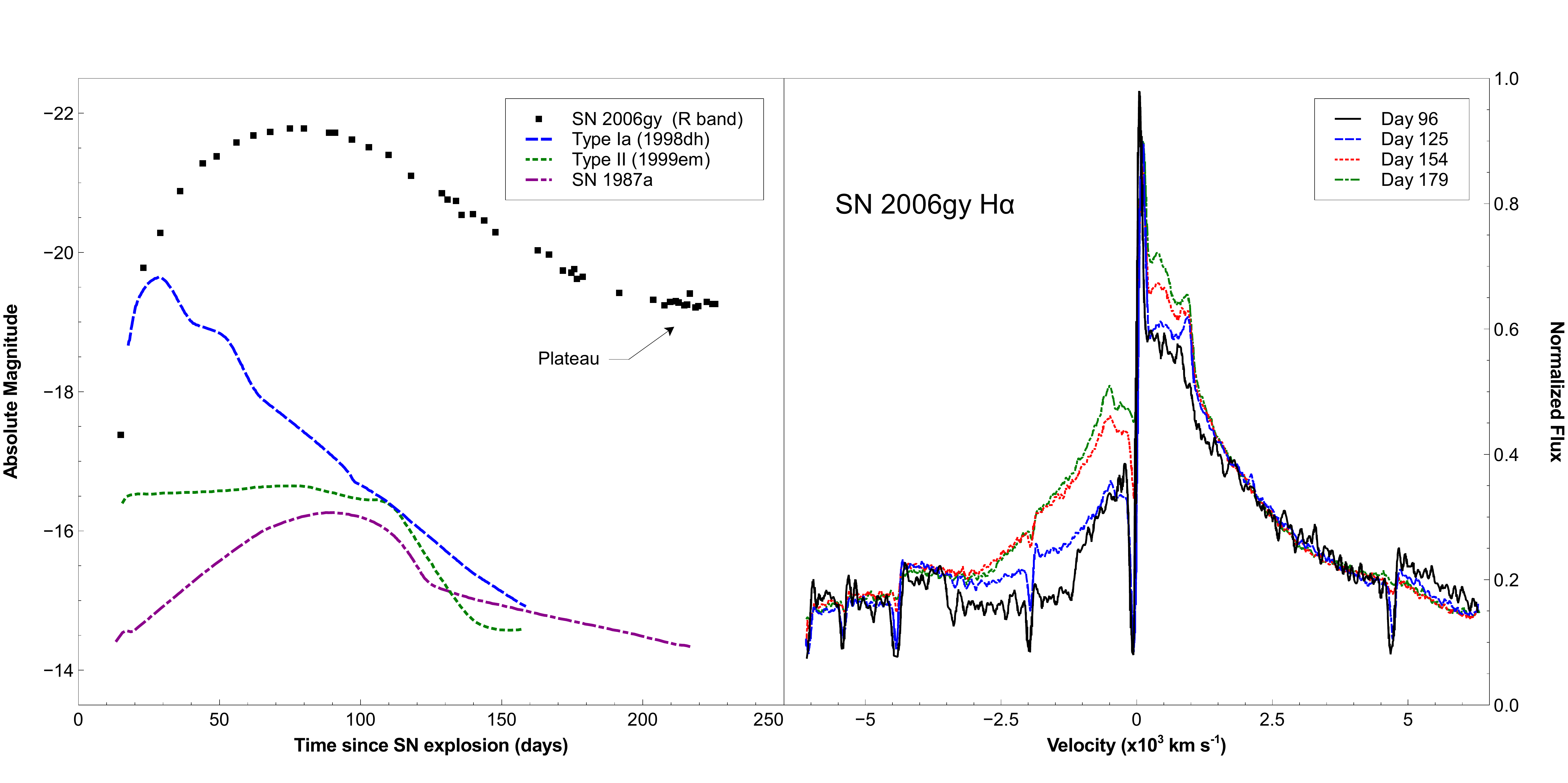}}
\caption{{\bf Photometric and spectroscopic observations of SN 2006gy. }
\textit{Left:}  The light curve (R-band) of SN 2006gy (black squares) is compared to light curves of three supernovae; a typical type Ia (SN 1998dh, blue long dash line), a typical type II (SN 1999em, green short dash line) and SN 1987a (mauve dash-dot line).  SN 2006gy is $\sim10$ times brighter than a typical type Ia and $\sim100$ times brighter than a typical type II.  As well, the light curve displays an unusual plateau beginning at $\sim$ day 200. \textit{Right:} Spectral observations of the SN 2006gy H$\alpha$ line for days 96 (black solid line), 125 (blue long dash line), 154 (red short dash line) and 179 (green dash-dot line) after inferred explosion date are overplotted and normalized to the same continuum level.  In this figure and all subsequent spectroscopic figures, the inferred explosion date is that cited by Smith et al. 2010.  The H$\alpha$ line displays a persistent broad component and a blue-side absorption feature that decreases with time. 
}
\label{obs}
\end{figure}

\begin{figure}
\resizebox{\hsize}{!}{\includegraphics{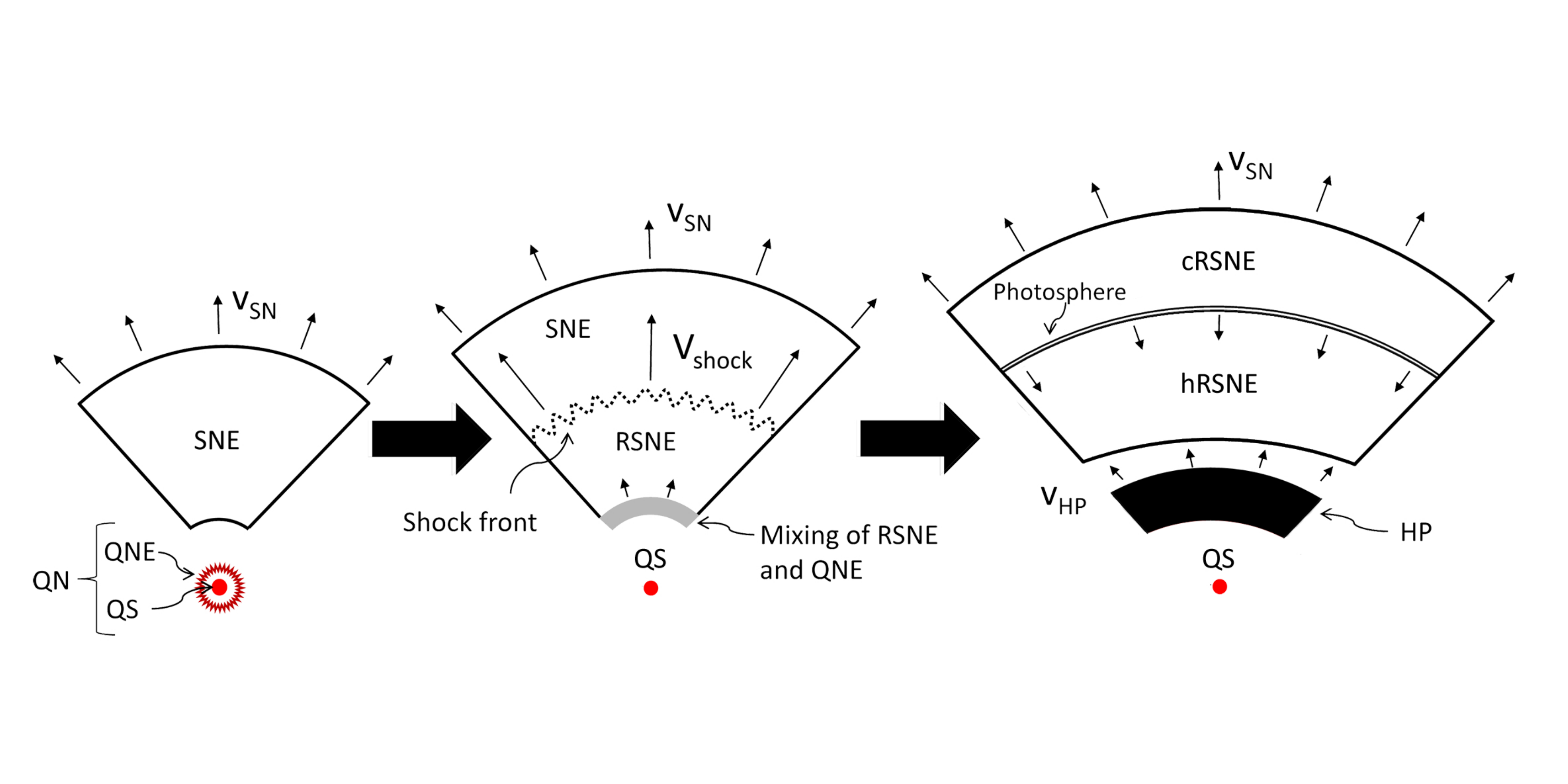}}
\caption{{\bf Schematic representation of the quark nova scenario. }
Displayed are cross sections of our model at three different phases. \textit{Left:} The instant the QN detonates inside the expanding envelope of SNE.  At this time the QNE is expelled at ultra-relativistic velocities leaving behind a quark star (QS).  \textit{Middle:} The interval in which the SNE is being shocked by the collision with the QNE.  Meanwhile at the inner edge of the envelope RSNE and QNE material are mixing to form the HP.  \textit{Right:} The evolution of the QN scenario after shock breakout.  Radiative cooling builds a cold outer layer of the RSNE (cRSNE) as the photosphere recedes inward.  At this time the HP is fully formed and coasting inside the RSNE.   
}
\label{schematic}
\end{figure}

\begin{figure}
\resizebox{\hsize}{!}{\includegraphics{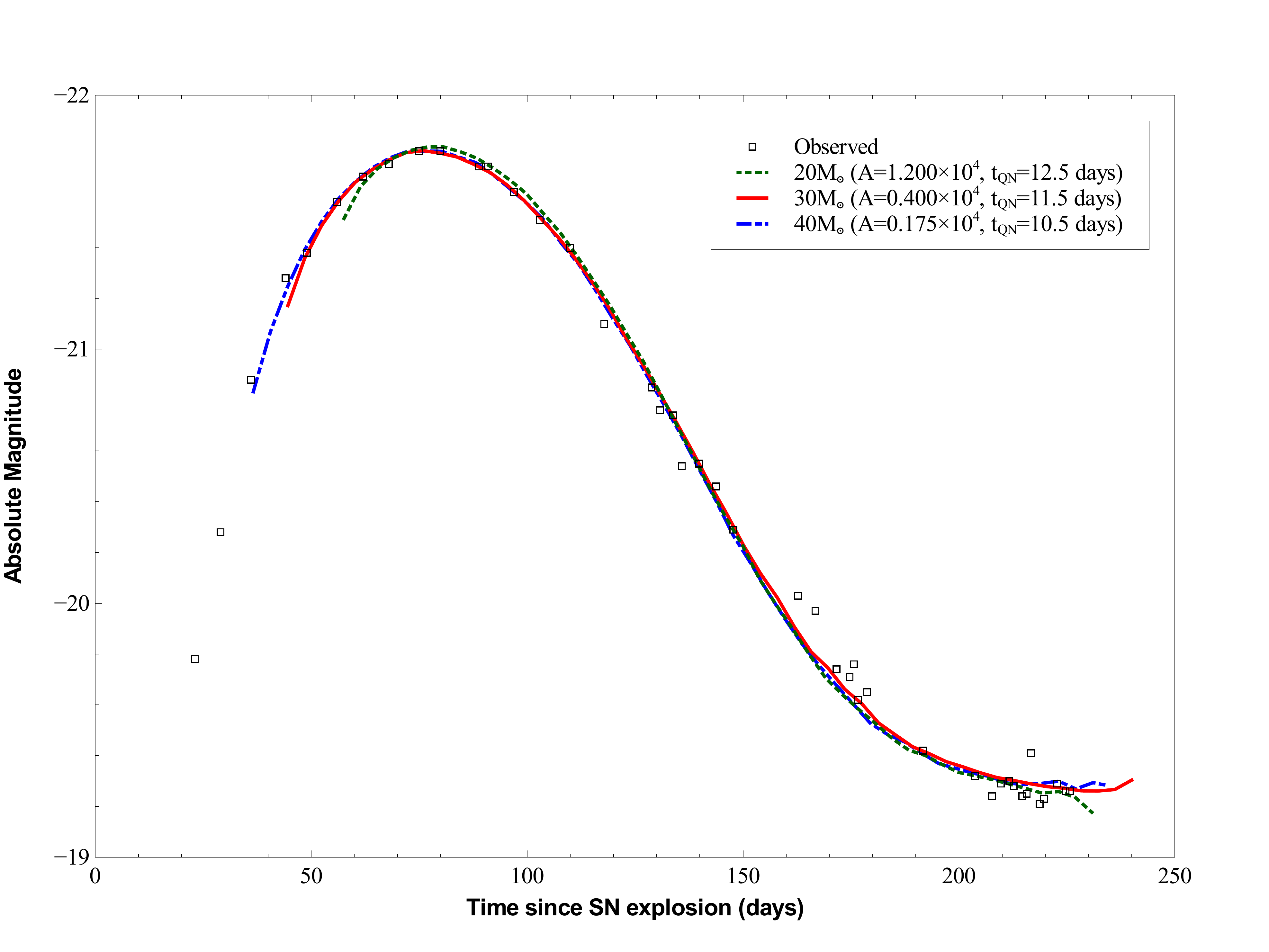}}
\caption{{\bf Comparison of models with different progenitor mass.}
The blue dash-dot line shows the R-band light curve fit for $M_t=40M_{\odot}$, the red solid line for $M_t=30M_{\odot}$, and the green dashed line for $M_t=20M_{\odot}$.  The parameters for each curve are given in table \ref{table:parameters}.  The open black squares represent R-band observations of SN 2006gy over 230 days beginning at the inferred explosion date of \citet{smith07}.  The different starting times for each curve reflect the different shock breakout times which is later for longer time delays. 
}
\label{figure:ms}
\end{figure}
 
\begin{figure}
\resizebox{\hsize}{!}{\includegraphics{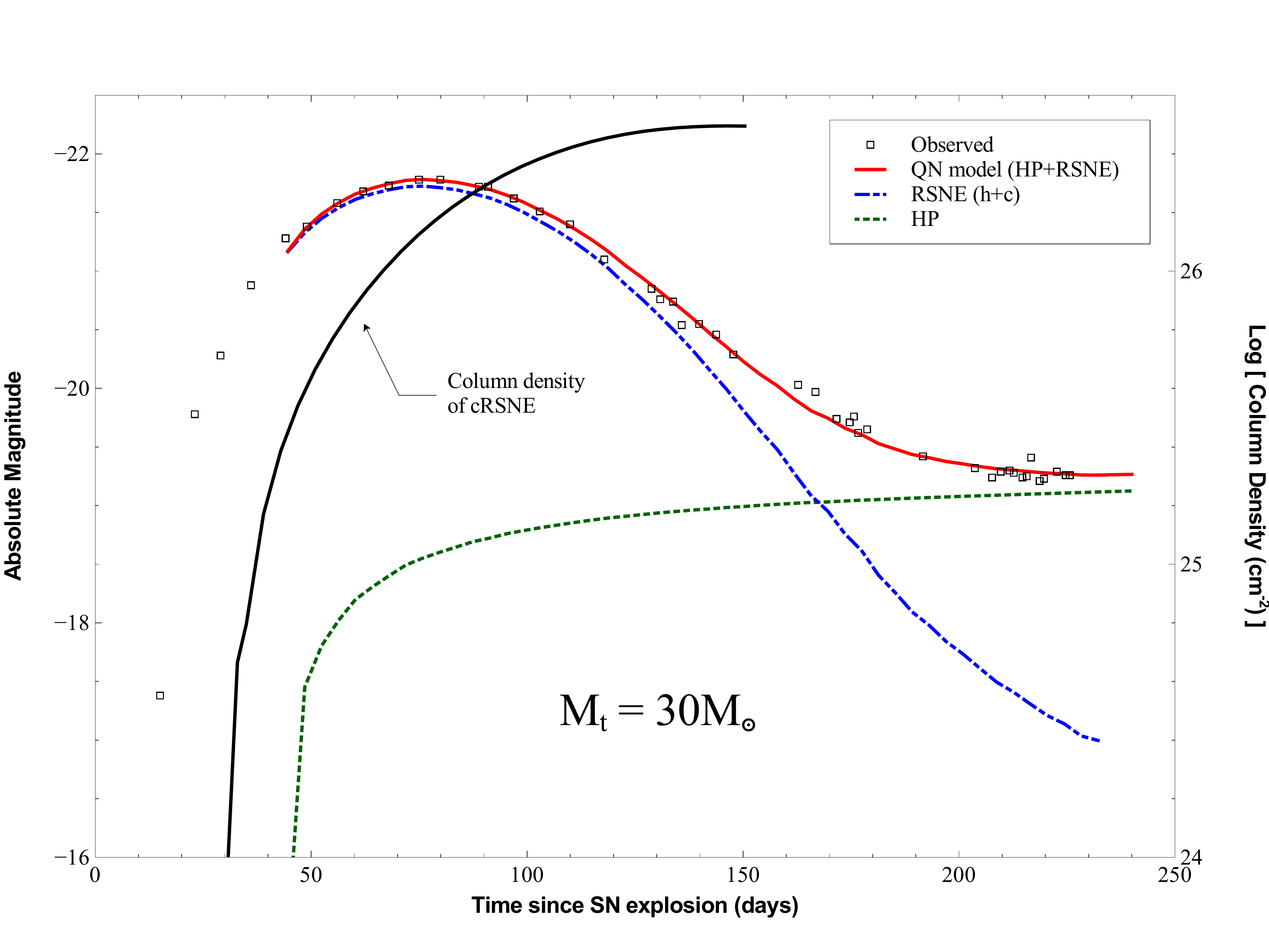}}
\caption{{\bf Comparison of the QN model to the R-band light curve of SN 2006gy. }
The red solid line shows the 30 $M_{\odot}$ best fit R-band light curve for our QN scenario which includes the RSNE and the HP.  Also plotted are the individual RSNE (blue dash-dot line) and HP (green dashed line) component R-band light curves as well as the column density of the cold outer layer of the RSNE (cRSNE) (black solid line). The open black squares represent the R-band observations of SN 2006gy over 230 days beginning at the inferred explosion date of \citet{smith07}.
}
\label{lcfit}
\end{figure} 
 
\begin{figure}
\resizebox{\hsize}{!}{\includegraphics{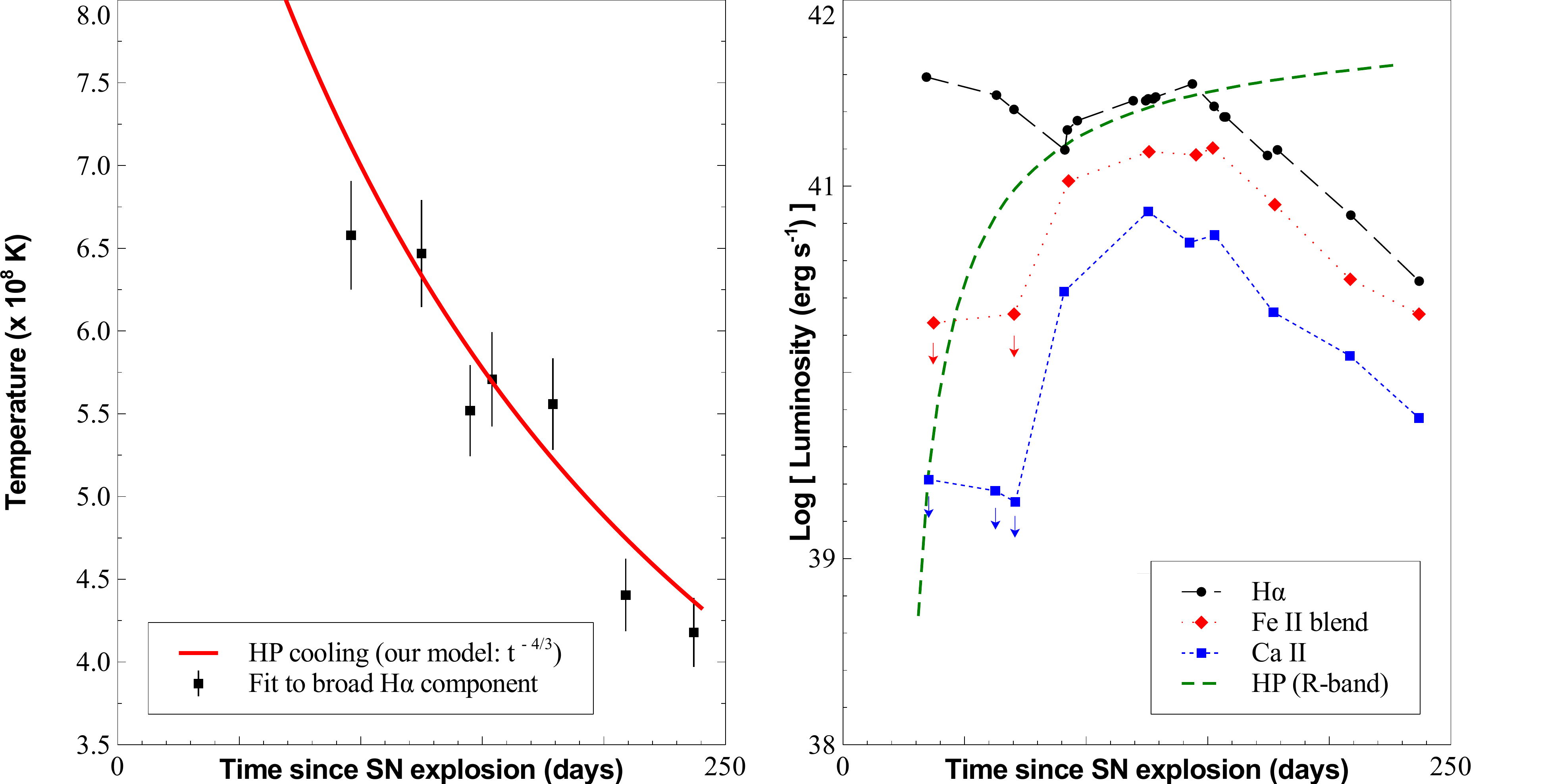}}
\caption{{\bf Further signatures of the HP. }
\textit{Left:} In this figure our predicted temperature of the HP is compared with values determined from fits to spectral observations of SN 2006gy.  The red solid line represents the temperature of the HP from our best fit light curve ($30M_{\odot}$).  Plotted as black closed circles are the HP temperatures found by fitting a Gaussian to the broad component of the H$\alpha$ line.  The error bars represent a 5 \% measurement error.
 \textit{Right:} This figure shows the evolution of the luminosities for the H$\alpha$ line, Ca II line at 8662 Angstroms and the red Fe II blend over roughly 7130-7628 Angstroms from Smith et al 2010.  After day 90 the Fe II and Ca II emission track the sudden peak in H$\alpha$ luminosity almost exactly.  This corresponds to the time when the opacity of the RSNE has dropped enough such that HP emission becomes a strong contributor to the overall light curve of SN 2006gy.  For comparison the R-band HP luminosity from our model light curve is plotted as a green dotted line.
}
\label{HPprop}
\end{figure}
 
\begin{figure}
\resizebox{\hsize}{!}{\includegraphics{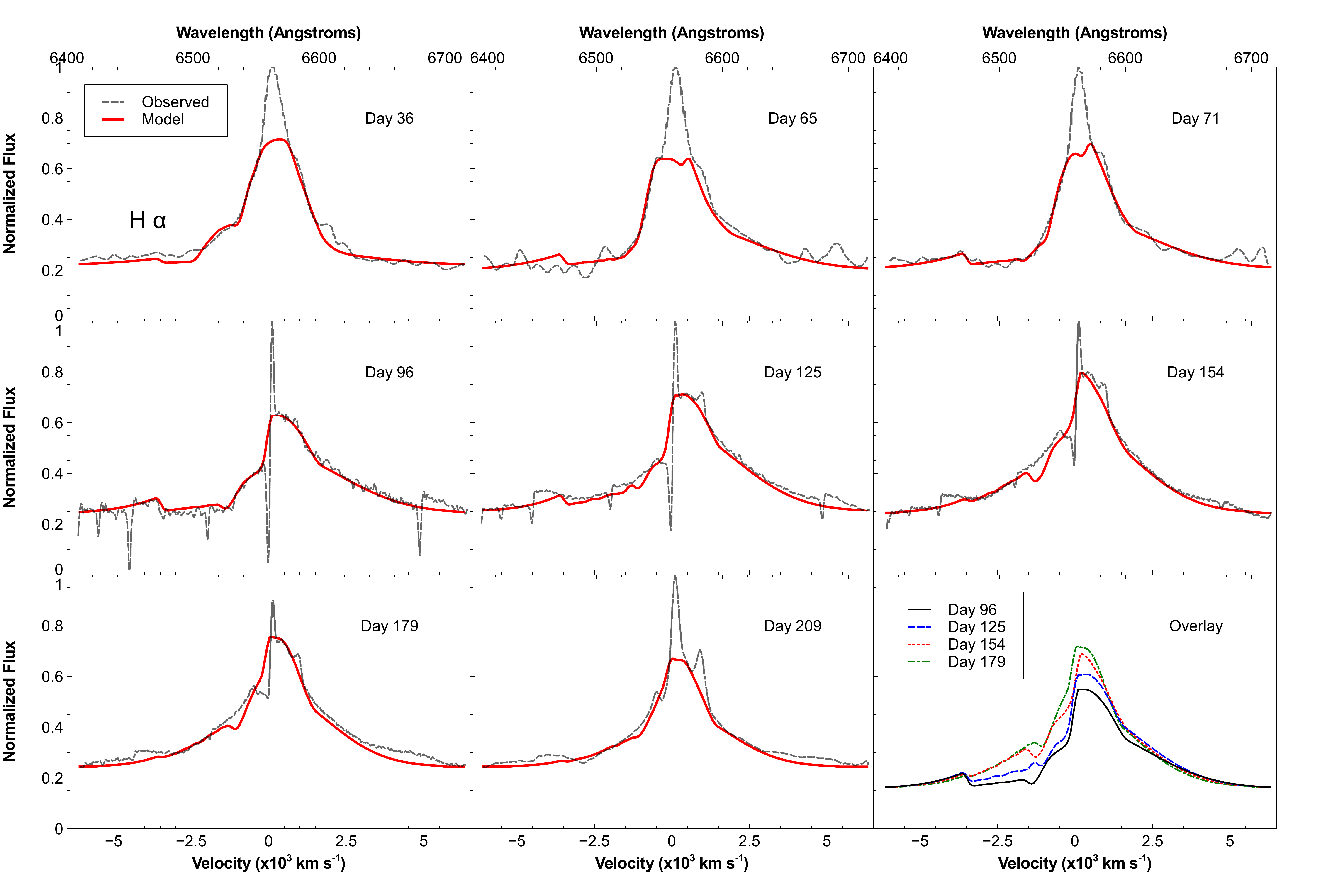}}
\caption{{\bf Evolution of the H$\alpha$ line in SN 2006gy.}
  This series displays the evolution of the observed H$\alpha$ line (grey dashed line) from SN 2006gy (data from Smith et al. 2010) compared with our model H$\alpha$ line (red solid line).  The dates shown are: 36, 65, 71, 96, 125, 154, 179 and 209 days after the inferred explosion date (\textit{left to right, top to bottom}).  The narrow peak in our model H$\alpha$ line is due to emission from the hRSNE.  The blue-side absorption feature is caused by the entire RSNE which has an outer edge velocity of 4000 km s$^{-1}$.  The broad component of the H$\alpha$ line is emission from the  HP which is coasting at a velocity of 95 km s$^{-1}$ and thus slowly cooling from its initial shock temperature ($0.9 \times 10^9$ K).  As the H$\alpha$ line evolves, the component due to the HP becomes more prominent and the blue-side absorption feature diminishes.  
 Displayed in the bottom right panel is our model H$\alpha$ line produced on days: 96, 125, 154 and 179 normalized to the same continuum level and overplotted.  The blue-side of the broad component shows strong absorption that decreases with time due to decreasing RSNE temperature and density, while the red-side remains notably constant.
}
\label{HAfit}
\end{figure}
 
\begin{figure}
\resizebox{\hsize}{!}{\includegraphics{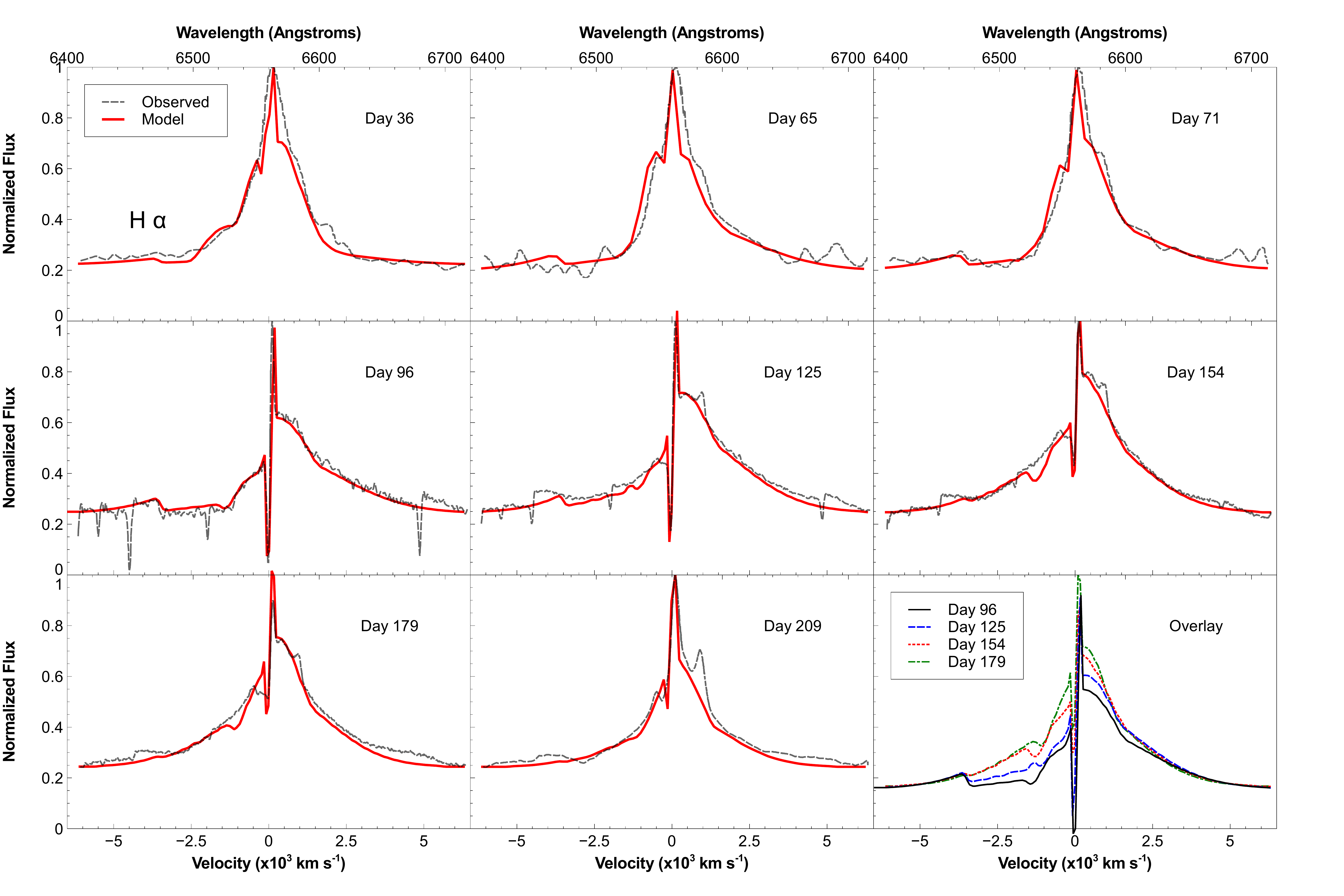}}
\caption{{\bf Evolution of the H$\alpha$ line in SN 2006gy including stellar wind.}
  This series displays the evolution of the observed H$\alpha$ line (grey dashed line) from SN 2006gy (data from Smith et al. 2010) compared with the H$\alpha$ line (red solid line) from our model which includes stellar wind from the progenitor.  The dates shown are:  36, 65, 71, 96, 125, 154, 179 and 209 days after the inferred explosion date (\textit{left to right, top to bottom}).  The intensity of the continuum emission from the wind is initially too low to allow for absorption.  During intermediate times the stellar wind continuum has increased such that a P-Cygni profile occurs, which is most prominent at day 96 and diminishes there after.  At late stages the stellar wind continuum is once again too low to allow for absorption. 
}
\label{wind}
\end{figure}

\begin{figure}
\resizebox{\hsize}{!}{\includegraphics{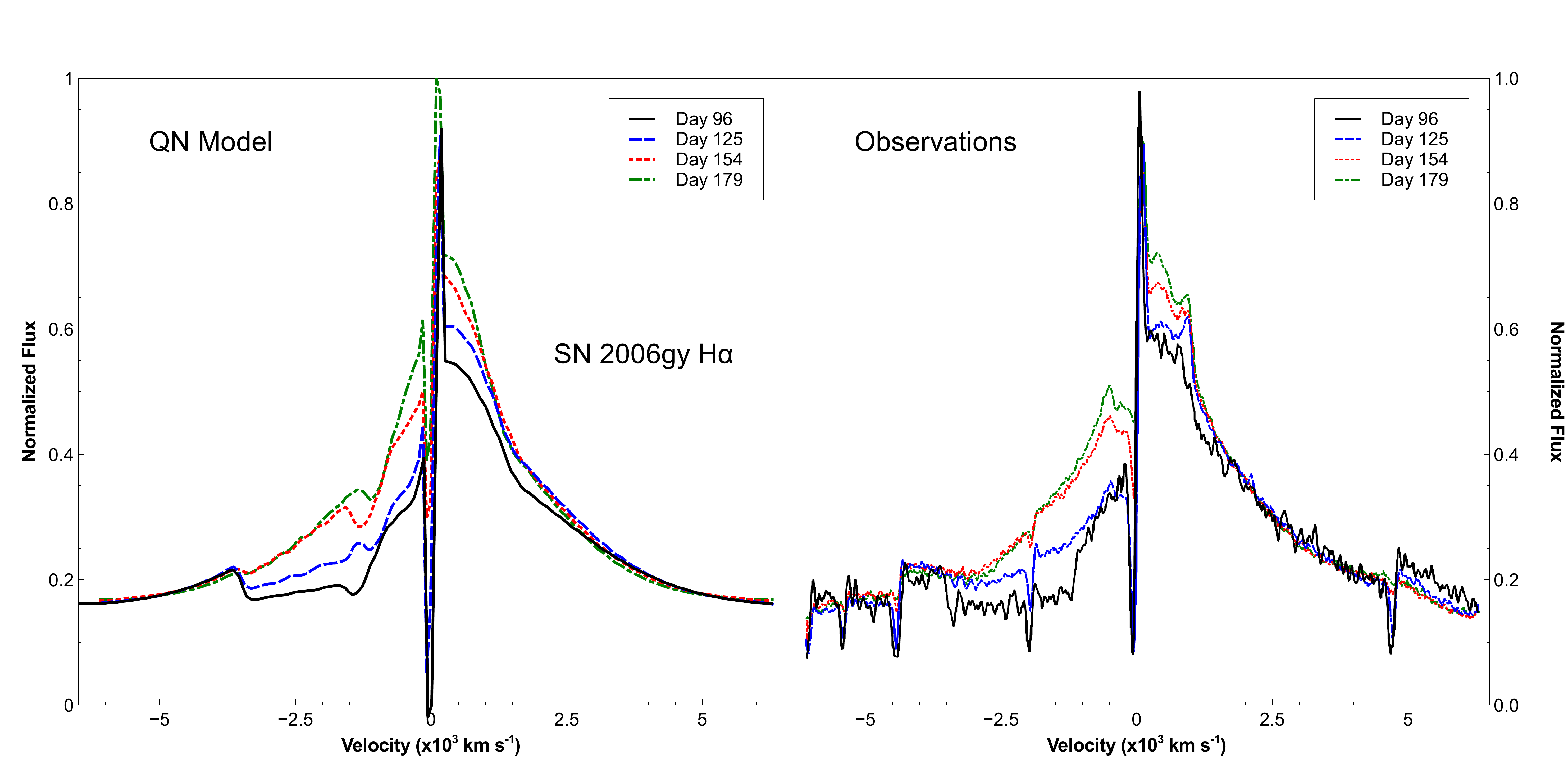}}
\caption{{\bf H$\alpha$ line over-plotted.}
For both panels of this figure the H$\alpha$ line from days 96, 125, 154 and 179 are normalized to the same continuum level and over-plotted. \textit{Left:} Displays the spectral line produced by our model. \textit{Right:} Displays the observations of SN 2006gy (data from Smith et al. 2010).
}
\label{Hoverlay}
\end{figure}

\begin{figure}
\resizebox{\hsize}{!}{\includegraphics{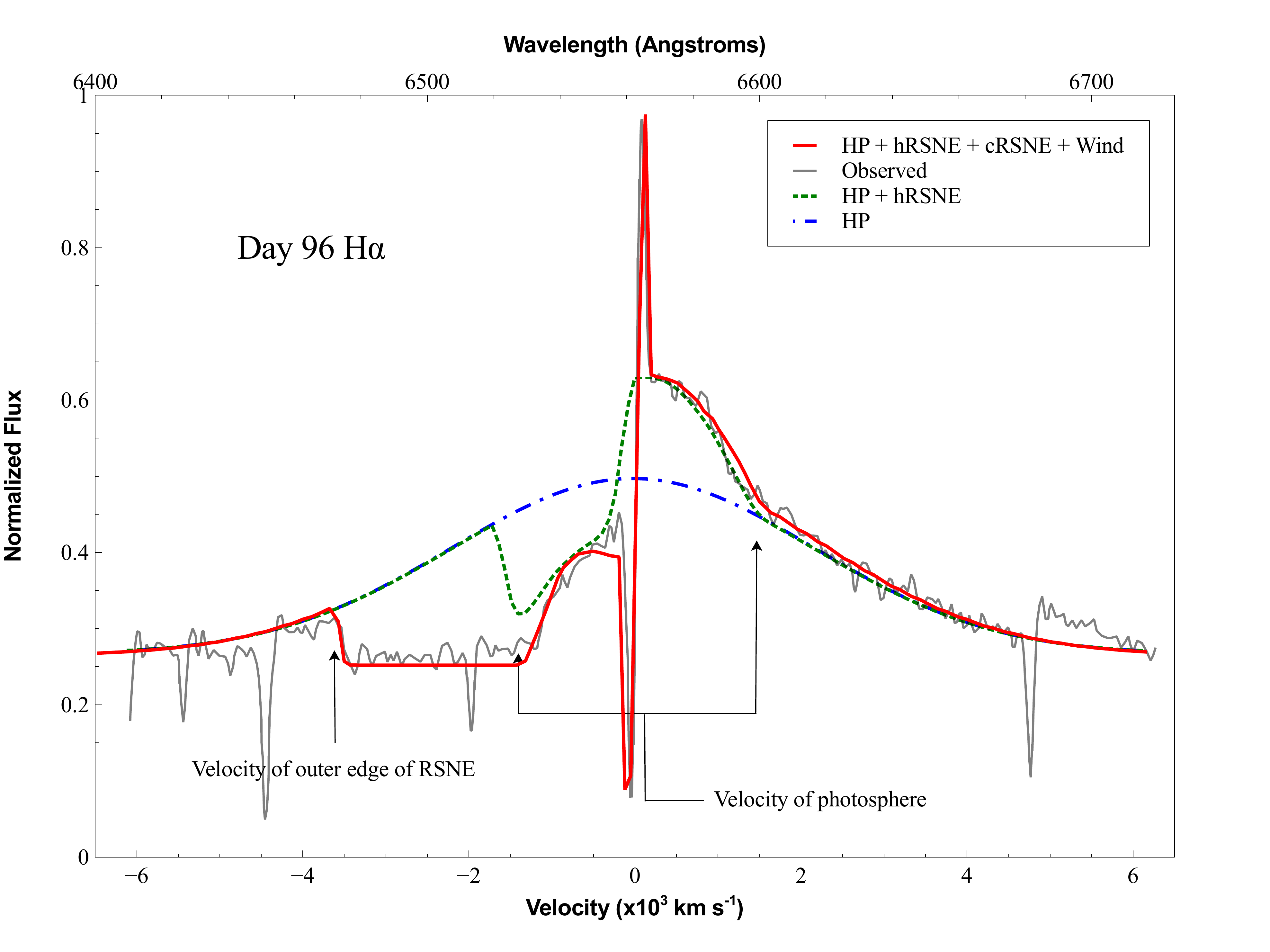}}
\caption{{\bf Features of the H$\alpha$ line.}
This figure shows the observed H$\alpha$ spectrum for day 96 (grey solid line) compared with the unabsorbed broad component from the HP (blue dash-dotted line) as well as our model H$\alpha$ line (red solid line) and a version that does not include the cold outer layer of the RSNE (cRSNE) (green dashed line).  The velocity of the outer edge of the RSNE and the RSNE photosphere are indicated with arrows.  This figure clearly demonstrates the contribution of our three components (HP, hRSNE and cRSNE) to the structure of the H$\alpha$ line.  
}
\label{d96}
\end{figure}

\end{onecolumn}

\end{document}